\documentclass[11pt, a4paper, final, twoside]{article}
\usepackage[centertags]{amsmath}
\usepackage{amssymb,amsmath}
\usepackage{amsfonts}
\usepackage{amssymb}
\usepackage{amsthm}
\usepackage{newlfont}
\usepackage{fancyhdr}
\usepackage{amscd}
\usepackage{sectsty}
\usepackage{siunitx}
\usepackage{graphicx}
\usepackage{subfigure}
\usepackage{cite}
\usepackage{etoolbox}
\usepackage{adjustbox}
\usepackage{booktabs}
\usepackage{array}
\usepackage{booktabs}
\usepackage[table]{xcolor}
\usepackage{xcolor,colortbl}
\usepackage{colortbl,multirow,hhline}
\usepackage{afterpage}
\definecolor{blue(ncs)}{rgb}{0.0, 0.53, 0.74}
\definecolor{blizzardblue}{rgb}{0.80, 0.91, 0.90}
\usepackage[colorlinks=true, urlcolor=blue(ncs), linkcolor=blue(ncs), citecolor=blue(ncs)]{hyperref}
\usepackage{caption}
\usepackage{latexsym,epstopdf,float}
\usepackage{euscript,pifont,mathrsfs,latexsym,graphicx}
\usepackage{mathrsfs}
\usepackage{etoolbox}
\usepackage[atend]{bookmark}

\bookmarksetup{
  open,
  openlevel=2,
  addtohook={%
    \ifnum\bookmarkget{level}>1 %
      \DisableBookmarkNumbering
    \fi
  },
}

\makeatletter
\newcommand*{\DisableBookmarkNumbering}{%
  \let\numberline\@gobble
}
\makeatother

\makeatletter
\renewcommand\@seccntformat[1]{}
\makeatother

\newcommand{\midsepremove}{\aboverulesep = 0.31mm \belowrulesep = 0.31mm}\midsepremove
\newcommand{\midsepdefault}{\aboverulesep = 0.605mm \belowrulesep = 0.984mm}\midsepdefault

\sectionfont{\color{blue(ncs)}}
\subsectionfont{\color{blue(ncs)}}
\patchcmd\thebibliography{\labelsep}{\labelsep\itemsep=0pt\parsep=0pt\relax}{}{\typeout{Couldn't patch the command}}

\renewenvironment{abstract}
{
\begin{center}%
\end{center}%
{\color{blue(ncs)}\normalfont\textbf{ABSTRCT:}}
}

\numberwithin{equation}{section}

\begin{document}
\pdfbookmark[0]{Unveiling the Complexity of Neural Populations: Evaluating the Validity and Limitations of the Wilson-Cowan Model}{frontmatter}
\title{Unveiling the Complexity of Neural Populations: Evaluating the Validity and Limitations of the Wilson-Cowan Model}

\author{
Maryam Saadati$^{1}$, Saba Sadat Khodaei$^{2}$, Yousef Jamali$^{2^{*}}$
}
\date{%
    $^1$ Department of Pure Mathematics, School of Sciences, Imam Khomeini International University, Qazvin, Iran \\
    $^2$ Biomathematics Laboratory, Department of Applied Mathematics, School of Mathematical Sciences, Tarbiat Modares University, Tehran, Iran \\
    $^*$ Corresponding author. Email: y.jamali@modares.ac.ir \\[2ex]%
}

\maketitle
\pdfbookmark[section]{ABSTRACT}{abstract}
\begin{abstract}
The population model of Wilson-Cowan is perhaps the most popular in the history of computational neuroscience. It embraces the nonlinear mean-field dynamics of excitatory and inhibitory neuronal populations provided via a temporal coarse-graining technique. The traditional Wilson-Cowan equations exhibit either steady-state regimes or else limit cycle competitions for an appropriate range of parameters. Since these equations set the coupled neural system at a lower resolution obscuring some potentially vital information, we focus on assessing the validity of approximations of the mass type model to a comprehensive description of complex neural behaviours. Starting from a large-scale network of threshold Hodgkin-Huxley style neurons, we formulate average population dynamics implicitly following from the mean-field assumptions. Our comparison of the microscopic neural activity with the macroscopic temporal profiles reveals dependency on the binary state of interacting subpopulations and the random property of structural network at the Hopf bifurcation points, when different synaptic weights are considered. For substantial configurations of stimulus intensity, our model provides further estimates of the neural population's dynamics official, ranging from simple periodic to quasi-periodic and aperiodic patterns, as well as phase transition regimes. While this shows its great potential for studying the collective behaviour of individual neurons with particularly concentrating on the occurrence of bifurcation phenomena, we must accept a quite limited accuracy of the Wilson-Cowan approximations-at least in some parameter regimes. Additionally, we report that the complexity and temporal diversity of neural dynamics, especially in terms of limit cycle trajectory, and synchronization can be induced by either small heterogeneity in the degree of various types of local excitatory connectivity or considerable diversity in the external drive to the excitatory pool.
\end{abstract}
\textbf{Keywords}\hspace{3mm}Wilson-Cowan model . Mean-field approximation . Graph theory . Hodgkin-Huxley model . Nonlinear dynamical system . Hopf bifurcation
\afterpage{} \fancyhead{} \fancyfoot{} \fancyhead[LE, RO]{\bf\thepage}
\section{INTRODUCTION}
\hypertarget{Intro}{}

Mathematical modeling of neuronal behaviour successfully covers multiple spatiotemporal scales, from single spiking neurons to large-scale networks of interacting brain regions \cite{Gelder(1988)}. At the microscale level, often associated with electrode recordings of single cell in vitro preparations, the Hodgkin-Huxley model and related conductance-based models are employed to explain the action potential properties of an individual neuron by introducing the nonlinear dynamics of multiple ion channels in coupled differential equations \cite{Hodgkin(1952),Hodgkin(1990),Rinzel(1990)}. At the opposite end, typically associated with functional imaging data, modeling the whole dynamics within brain areas fairly contributes to interpreting the activity of state variables and advancing our understanding of the dynamical organization of the brain \cite{Vertes(2012)}.

The middle of these two ends consists of well-known population-level models including \textit{Neural-mass} models \cite{Wilson(1972),Wilson(1973),Lopes(1974),Freeman(1975),Jansen(1995),David(2003),Stefanescu(2008),Ponten(2010),Byrne(2017),Byrne(2019)} and \textit{Neural-field} models \cite{Nunez(1995),Jirsa(1996),Amari(1977),Bressloff(2011),Coombes(2014),Bressloff(2019)}. These types of models employ coarse-grained and mean-field techniques, first employed by Amari \cite{Amari(1972)}, to characterize the collective dynamics typically measured via local-field potentials (LFP), electroencephalography (EEG), and magnetoencephalography (MEG) \cite{Freeman(1987),Deco(2008),Kozma(2016)}. Coarse-graining considers the neural system at a lower dimension, leading to low degree of freedom and loss of some information in the dynamical system. Mean-field theory initially proposed in magnetics \cite{Bellac(2004),Peliti(2011)} captures the relevant mean activity at spatiotemporally coarse-grained scales while neglecting correlations and fluctuations around the mean.

Among the phenomenological neural-mass models, the Wilson-Cowan type model is perhaps the most popular. The model, developed in 1972 by Wilson and Cowan \cite{Wilson(1972)}, provides a coarse-grained approximation of firing activities of two interacting populations of neurons--one of which is excitatory (E) and the other inhibitory (I). They were inspired by physiological evidence from Mountcastle \cite{Mountcastle(1957)} and Hubel and Wiesel \cite{Wiesel(1963),Hubel(1965)} and anatomical evidence from Colonnier \cite{Colonnier(1965)} and Szent\'{a}gothai and Lissak \cite{Szentagothai(1967)}, which supported the existence of certain populations of neurons with very nearly identical responses to identical stimuli; for a nice historical perspective on the development of their ideas, we refer the interested reader to \cite{Chow(2020),Destexhe(2009),Anderson(1998)}. Building on earlier works by Sholl \cite{Sholl(1955)} and Beurle \cite{Beurle(1956)}, Wilson and Cowan claimed that a low-level (single neuron) description of qualitative behaviour is not probably suited for understanding higher-dimensional functions such as sensory processing, memory storage, learning, and pattern recognition. Using the techniques from phase plane analysis and numerical bifurcation theory, they showed that their equations exhibit switching and cycling, as well as multistability and hysteresis which could be perceived as a basis for information storage \cite{Cragg(1955),Fender(1967)}.

The original Wilson-Cowan framework and variants, as a dynamical description for a single EI population, a building unit for large-scale brain networks, and a basis for spatially extended models at the tissue scale, have now been highly impactful in explaining a substantial range of observed cortical phenomena such as visual hallucinations \cite{Ermentrout(1979),Bressloff(2001),Rule(2011),Bertalmio(2020)}, binocular rivalry \cite{Wilson(2001),Lee(2007),Bressloff(2012),Wilson(2017)}, epilepsy \cite{Shusterman(2008),Meijer(2015)}, Parkinson's disease \cite{Kaslik(2021)}, patient tremor \cite{Duchet(2020)}, traveling cortical waves \cite{Wilson(2001),Roberts(2019)}, conscious decision interpreted as a form of generalized rivalry among several plausible alternative possibilities \cite{Wilson(2009),Wilson(2013)}, cognitive dynamics of movement \cite{Erlhagen(2002)}, phase-amplitude coupling \cite{Onslow(2014)}, and cortical resonant frequencies \cite{LeaCarnall(2016)}, just to name a few. For more applications of the Wilson-Cowan model, we refer the interested reader to the following works of Coombes \cite{Coombes(2005)} and Wilson and Cowan \cite{Wilson(2021)}.

While the population-level model of Wilson-Cowan is able to produce different dynamical modes of activity, depending on a few parameters, it is unconstrained by the biophysical underpinnings that can be important to forecast the final outcome of electrical or pharmacological stimulation in experimental or clinical conditions \cite{Anticevic(2012),Murray(2014)}. The Hodgkin-Huxley type model, on the other hand, is a biophysical model considering detailed molecular mechanisms that govern ionic currents into biologically plausible parameters at the neuronal level. Here, we combine the two models to retain both the men-field restrictions of the Wilson-Cowan model and the biophysical constraints of the Hodgkin-Huxley model. The unified nonlinear-dynamical model is thus appropriate for questioning the extent to which neural-mass models really resemble the mean dynamics of localized EI population and for studying the effect of physiological and structural properties on the temporal diversity of neural behaviour. We show that near the boundaries at which transitions occur, the complexity of dynamical landscape is determined by both the binary operation inherited from the Hodgkin-Huxley style population and the randomness degree considered in the connection topology, however, cannot be predicted based on the initial macroscale information. Specifically focusing on the bifurcation regions, the complete characterization of quasi-periodic and complex chaotic states expected from a neural model are obtained for the microscopic network of neurons, which is a major difference with the Wilson-Cowan model in its most simple incarnation. Together, this work shows how the topological properties and the external currents of several classes can affect the behavioural dynamics of microscale neurons to establish a desired relation between temporal response to diversity in the range of stimulation and in the degree of excitatory type coupling.
\begin{figure}[!t]
\hypertarget{Fig1}{}
\centering
{
\includegraphics[width=\textwidth]{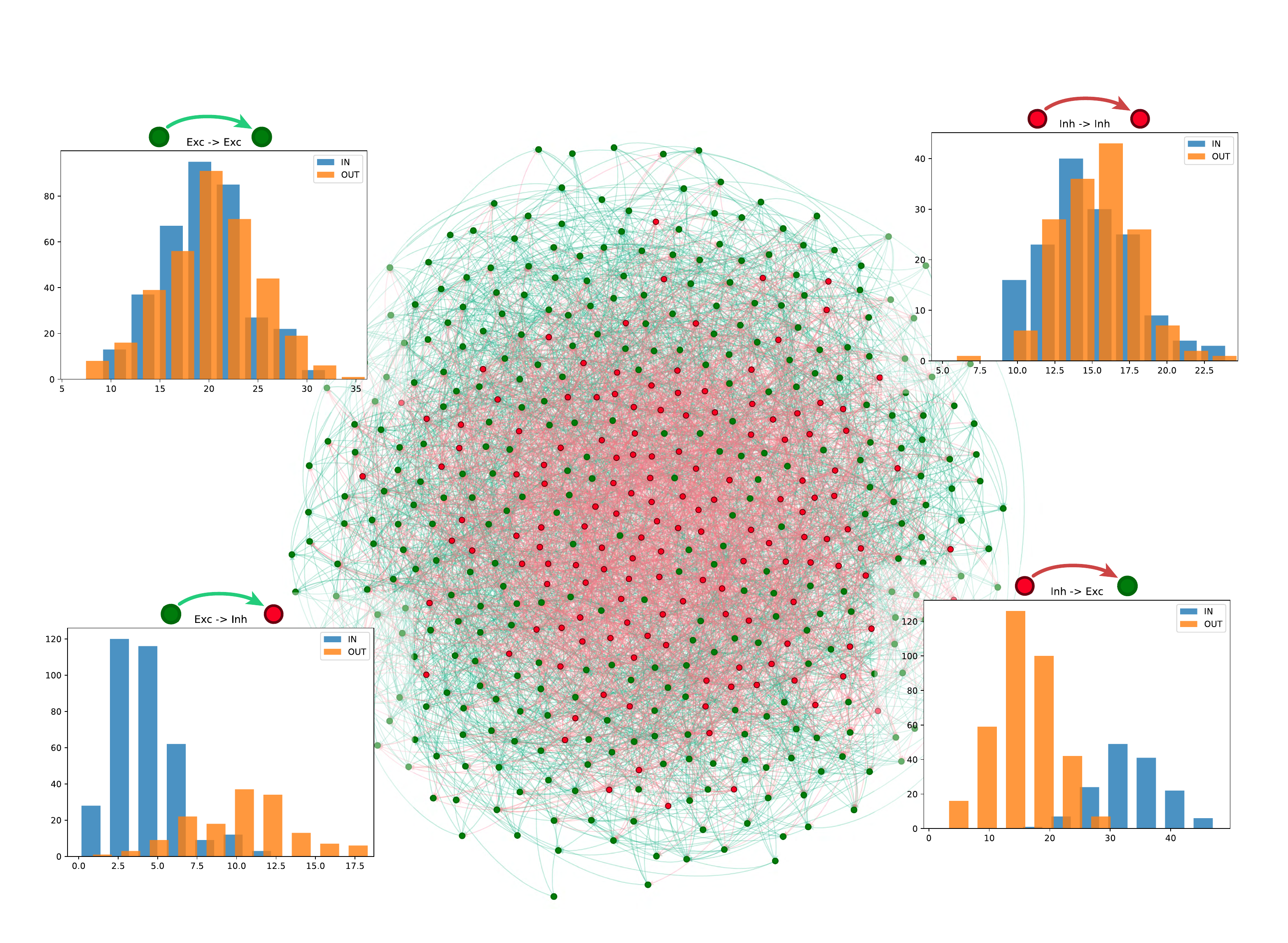}
}
\vspace{-7mm}
\caption{\footnotesize\textbf{A schematic representation of the Hodgkin-Huxley neural network under study.} Green or red circles represent excitatory or inhibitory neurons, respectively. Green arrows denote input or output excitatory connections as bidirectional and red arrows correspond to inhibitory ones.}
\end{figure}
\section{RESULTS}
\hypertarget{Res}{}
\subsection{Dynamical properties of bifurcations arising out of coupling strength in the microscopic system}
\hypertarget{Res1}{}

To evaluate the quality of the neural-mass theory's approximations, we propose a conductance-based neural network capturing the restrictions of a rate-based mass model, and accomplish the phase plane analysis to compare the impact of physiological parameters with their macroscale counterparts on the temporal behaviour of localized neuronal populations. As the first step, it raises a fundamental question how the synaptic connection weight close to bifurcation points can cause different dynamical regimes in the system with predefined microscopic neurons.

For a range of low coupling strengths, we observe that the excitatory population's firing converges to an almost constant rate with a low value despite receiving a sufficient stimulus, and that the inhibitory population settles into a resting state with zero activity, tuning the neural system at a low firing region with stable fixed points (Figure\,\hyperlink{Fig2}{2\,A}). Increasing the average synaptic weight, as shown in Figure\,\hyperlink{Fig2}{2\,B}, not only does the number of active neurons increase and local synchronization regimens enhance but also oscillatory patterns similar to those of the Wilson-Cowan oscillators appear in the output firing rates of micro populations, which in turn lead to a spontaneous switch between the steady-state solutions and Hopf bifurcation limit cycles at a periodic firing state. Indeed, the qualitative threshold for the limit cycle activity to emerge depends on the quantitative threshold of synaptic strength dynamics to exceed. And especially, the strength of internal connections is inversely related to both the frequency of limit cycle oscillations and the phase difference between the coupled excitatory and inhibitory oscillators, as well as directly related to the amplitude of this oscillatory response because of decreasing the neural activity level and the number of spikes per unit time (Figure\,\hyperlink{Fig2}{2\,D}). At higher synaptic weights, while being extremely sensitive to the external input, the system moves towards a noisy regular firing region with high frequency and a nearly constant, high-value rate, where the limit cycle trajectory sharply jumps to a high firing stable fixed point through a Hopf bifurcation (Figure\,\hyperlink{Fig2}{2\,E}).

Notice that the qualitative regulation effects of structural connectivity strength at the large-scale network level are roughly consistent with those of the neural-mass model via its Wilson-Cowan dynamics, but there are two remarkable differences between the output rate of these models at the microscale and macroscale. First, our EI population exhibits substantial binary modes at the low and high rate fixed points, despite the dynamical behaviour of the mean-field equations growing or shrinking to bifurcation points gradually, given Figures\,\hyperlink{Fig2}{2\,A}, \hyperlink{Fig2}{B}, \hyperlink{Fig2}{D}, and \hyperlink{Fig2}{E}. In fact, the binary-state dynamics of the Hodgkin-Huxley biophysical model induces active to inactive phase transitions in the Hopf bifurcations of the fixed points emerging from an increase of the coupling weight. Second, near the phase transition points, the random network topology neglected in the Wilson-Cowan assumptions strongly affects the microscopic dynamics, which causes the system to wander around these critical points and displays the up-down state transitions of membrane potential and population firing rates (Figures\,\hyperlink{Fig2}{2\,A}-\hyperlink{Fig2}{G}). In other words, concerning the stochastic nature of our neural dynamics, the two-dimensional position of bifurcation points is not precisely determined and the probability that limit cycle competitions occur depends on an operating interval. These discrepancies originate from the mean-field approximation of coarse-grained neural activity that deliberately neglects stochastic effects to make the macroscopic equations tractable. It is worth mentioning that the regular oscillations observed in the firing rate dynamics arise from both following cases: (i) the synchronization degree between neurons (ii) the time-dependent region between the activity and non-activity states. Actually, the bursting and quiescent behaviours defining the onset of synchrony in the EI network generate the oscillatory firing rate regimes; hence, the beginning of locally synchronized fashion at the single neuron level reflects making the limit cycle condition of the simplest Wilson-Cowan equations at the population level. Furthermore, the high fixed points are necessarily indicative of the high synchronous regimes, although having the approximately constant rates of activity like the stable fixed points close to low firing regimess.

Our results show that in the microscopic dynamics, the specific features of neural transitions controlled effectively by the synaptic strength do not follow the mean-field deterministic trajectory, suggesting that the phenomenologically derived Wilson-Cowan model is relatively far from reality.
\begin{figure}[!t]
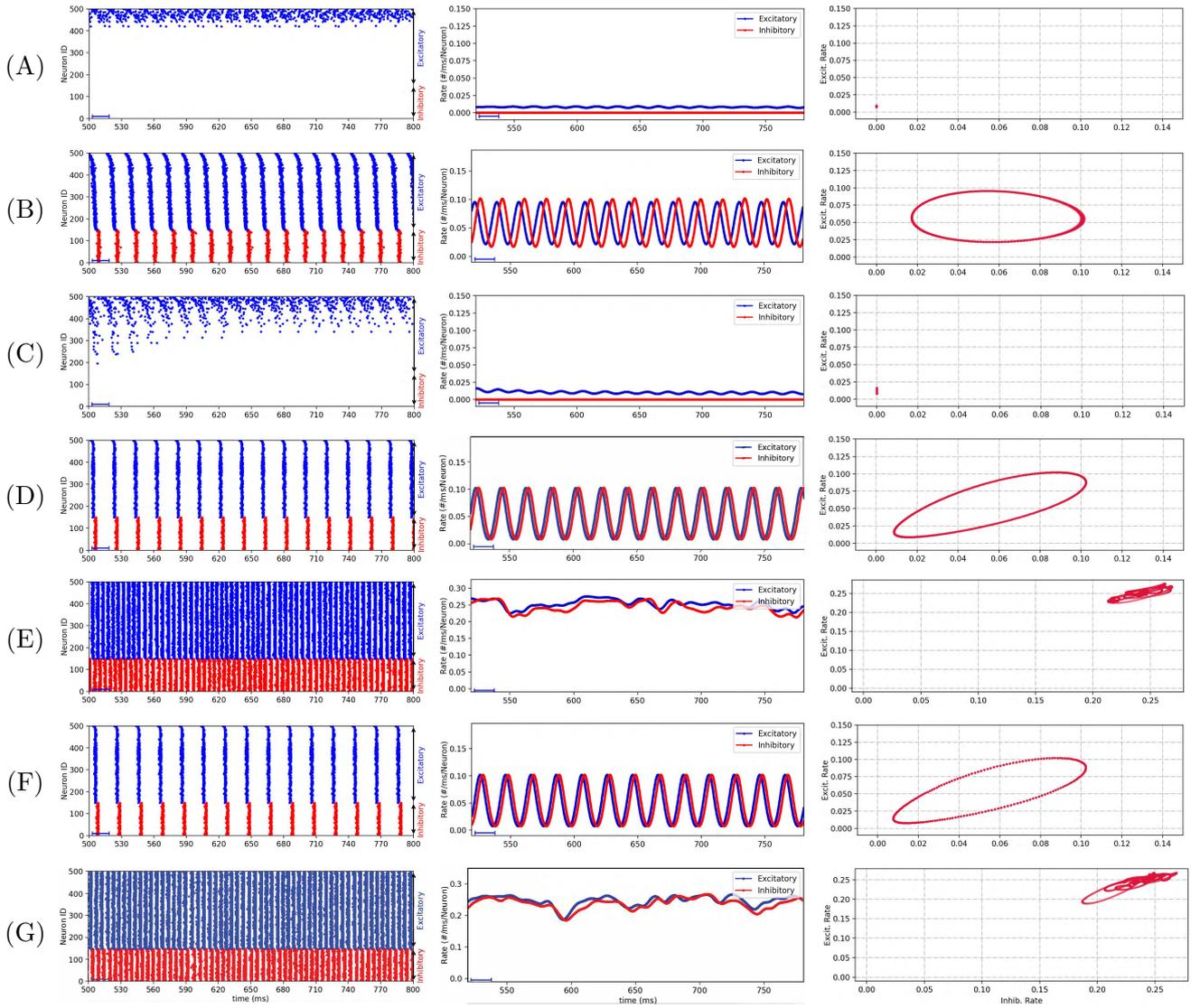

\hypertarget{Fig2}{}
\begin{center}
\begin{tabular}{@{\hspace*{0mm}}c
                @{\hspace*{2mm}}c
                @{\hspace*{2mm}}c
                @{\hspace*{2mm}}c}
(A)&
\adjustimage{width=0.31\textwidth, valign=c}{2-1-1}&
\adjustimage{width=0.31\textwidth, valign=c}{2-1-2}&
\adjustimage{width=0.31\textwidth, valign=c}{2-1-3}
\medskip
\\
(B)&
\adjustimage{width=0.31\textwidth, valign=c}{2-2-1}&
\adjustimage{width=0.31\textwidth, valign=c}{2-2-2}&
\adjustimage{width=0.31\textwidth, valign=c}{2-2-3}
\medskip
\\
(C)&
\adjustimage{width=0.31\textwidth, valign=c}{2-3-1}&
\adjustimage{width=0.31\textwidth, valign=c}{2-3-2}&
\adjustimage{width=0.31\textwidth, valign=c}{2-3-3}
\medskip
\\
(D)&
\adjustimage{width=0.31\textwidth, valign=c}{2-4-1}&
\adjustimage{width=0.31\textwidth, valign=c}{2-4-2}&
\adjustimage{width=0.31\textwidth, valign=c}{2-4-3}
\medskip
\\
(E)&
\adjustimage{width=0.31\textwidth, valign=c}{2-5-1}&
\adjustimage{width=0.31\textwidth, valign=c}{2-5-2}&
\adjustimage{width=0.31\textwidth, valign=c}{2-5-3}
\medskip
\\
(F)&
\adjustimage{width=0.31\textwidth, valign=c}{2-6-1}&
\adjustimage{width=0.31\textwidth, valign=c}{2-6-2}&
\adjustimage{width=0.31\textwidth, valign=c}{2-6-3}
\medskip
\\
(G)&
\adjustimage{width=0.31\textwidth, valign=c}{2-7-1}&
\adjustimage{width=0.31\textwidth, valign=c}{2-7-2}&
\adjustimage{width=0.31\textwidth, valign=c}{2-7-3}
\end{tabular}
\end{center}
\caption{\footnotesize\textbf{The network's dynamical characteristics close to bifurcation points over the global synaptic weight parameter $W$.} First column: Raster plots of the activity for $500$ Hodgkin-Huxley neurons sorted by input intensity. Second column: Evolution of state variables $Rate (t)$ of the EI system. Third column: Phase planes showing temporal trajectories in response to constant stimulation. The coupling strength drives the microscopic population to different binary modes in which the system wanders around the active-inactive phase transitions. The specific parameter defining the global synaptic weight is {\bf (A)} $W = 0.005$ \si{\siemens/\square{\centi\meter}}, {\bf (B)} $W = 0.00942$ \si{\siemens/\square{\centi\meter}}, {\bf (C)} $W = 0.00942$ \si{\siemens/\square{\centi\meter}}, {\bf (D)} $W = 0.5$ \si{\siemens/\square{\centi\meter}}, {\bf (E)} $W = 0.681$ \si{\siemens/\square{\centi\meter}}, {\bf (F)} $W = 0.683$ \si{\siemens/\square{\centi\meter}}, and {\bf (G)} $W = 0.683$ \si{\siemens/\square{\centi\meter}}.}
\end{figure}
\subsection{Chaos and periodic sequences in the interconnected EI network under different stimulus intensities}
\hypertarget{Res2}{}

In what follows, focusing on the occurrence of bifurcation phenomena, we study initially the role played by the excitatory pathway in bringing the micro populations to the distinct regimes of activity, rendering a comparison with the time series of the neural-mass model's outcome. Afterward, we perform the noise sensitivity with respect to parameters that take both the average coupling weight and the input variance profiles.

In the case of low input rate, corresponding to Figure\,\hyperlink{Fig3}{3\,A}, the output firing rates of local populations do not show exquisite sensitivity to the external drive, which in turn produce a stable quiescent fixed point of Hopf bifurcation. Sufficiently large stimulus levels set the receiving network close to the threshold of activation so that the microscopic system makes a sharp synchrony response to the excitatory current, whilst escaping from the low activity background state to a limit cycle trajectory with medium firing rates (Figure\,\hyperlink{Fig3}{3\,B}). This oscillatory region in the phase space is consistent with the imbalance dynamics between inhibition and excitation which leads to regular firing patterns at the individual neuron level as well as periodical behaviours resembling the Wilson-Cowan oscillations at the population level. According to Figure\,\hyperlink{Fig3}{3\,C}, when increasing the external rate further, on one hand, the EI population enters a more synchronous regime with higher spike frequency, and on the other hand, the activity level in the excitatory groups gets higher than in the inhibitory ones because of applying the driving force only to the excitatory subpopulation. In addition, both the frequency of neural population firing rates and the phase difference between their respective oscillations depend linearly on the input strength while their amplitude is inversely related to it which corresponds to shrinking the basin of attraction of the down-state fixed point. This might be attributed to an increase in the number of action potentials per unit time induced by an excess in the outside stimulation. Figure\,\hyperlink{Fig3}{3\,F} illustrates that, due to the hyperpolarizing state considered in the Hodgkin-Huxley modeling, the E population activity reaches a saturation level faster as the strength of excitatory drive greatly increases, leading to the annihilation of temporal oscillations in the phase space. Finally, significantly higher levels of the external noise kick the system in a region of high fixed point solutions with a stable profile where the population rate dynamics is very weakly dependent on the value of input current (Figure\,\hyperlink{Fig3}{3\,G}).

Analyzing and comparing to the dynamical behaviour of the original Wilson-Cowan reference, in a specific range of the input intensity, our simulations of the microscale neurons reveal quasi-periodic solutions and chaotic patterns related to synchronous complicated bursting in the $(\textrm{Excit. Rate}, \textrm{Inhib. Rate})$-phase space, which interestingly could not be observed in the case of mean-field activity equations (see Figures\,\hyperlink{Fig3}{3\,C}, \hyperlink{Fig3}{D}, and \hyperlink{Fig3}{E}). Moreover, in the vicinity of the low activity fixed point, the pulse response of self-interacting EI network to raising the external load is binary again, owing to the binary operation of the Hodgkin-Huxley neurons which evokes a phase transition at this Hopf bifurcation point (Figures\,\hyperlink{Fig3}{3\,A} and \hyperlink{Fig3}{B}). These essential differences emphasize that concentrating solely on the macroscopic neural-mass level, which averages out the population's spiking treatment, is inadequate for understanding and completely covering the complexity of intrinsic neural dynamics.

It is noteworthy that in the phase plane trajectories for the mentioned system, there exist regimes in which there is not any periodic solution consistent with the Wilson-Cowan population's behaviour, as depicted in Figure\,\hyperlink{Fig4}{4} and Figure\,\hyperlink{Fig5}{5}. While keeping the synapse's average weight down or the input variance up, which cause a lower degree of synchrony (e.g., compare Figure\,\hyperlink{Fig3}{3\,C}, Figure\,\hyperlink{Fig4}{4\,C}, and Figure\,\hyperlink{Fig5}{5\,C}), increasing the mean of the injected current to the excitatory pool compensates for the influence of these particular parameters so that the micro population of neurons tends to the synchronization more pronouncedly; however, the system operates only near the stable low and high fixed points. Indeed, an appropriate choice of the neural coupling strength plus the standard deviation in the input distribution can ensure the local EI network capable of producing limit cycle bifurcations arising from middle-range synchronous firing states.
\begin{figure}[!t]
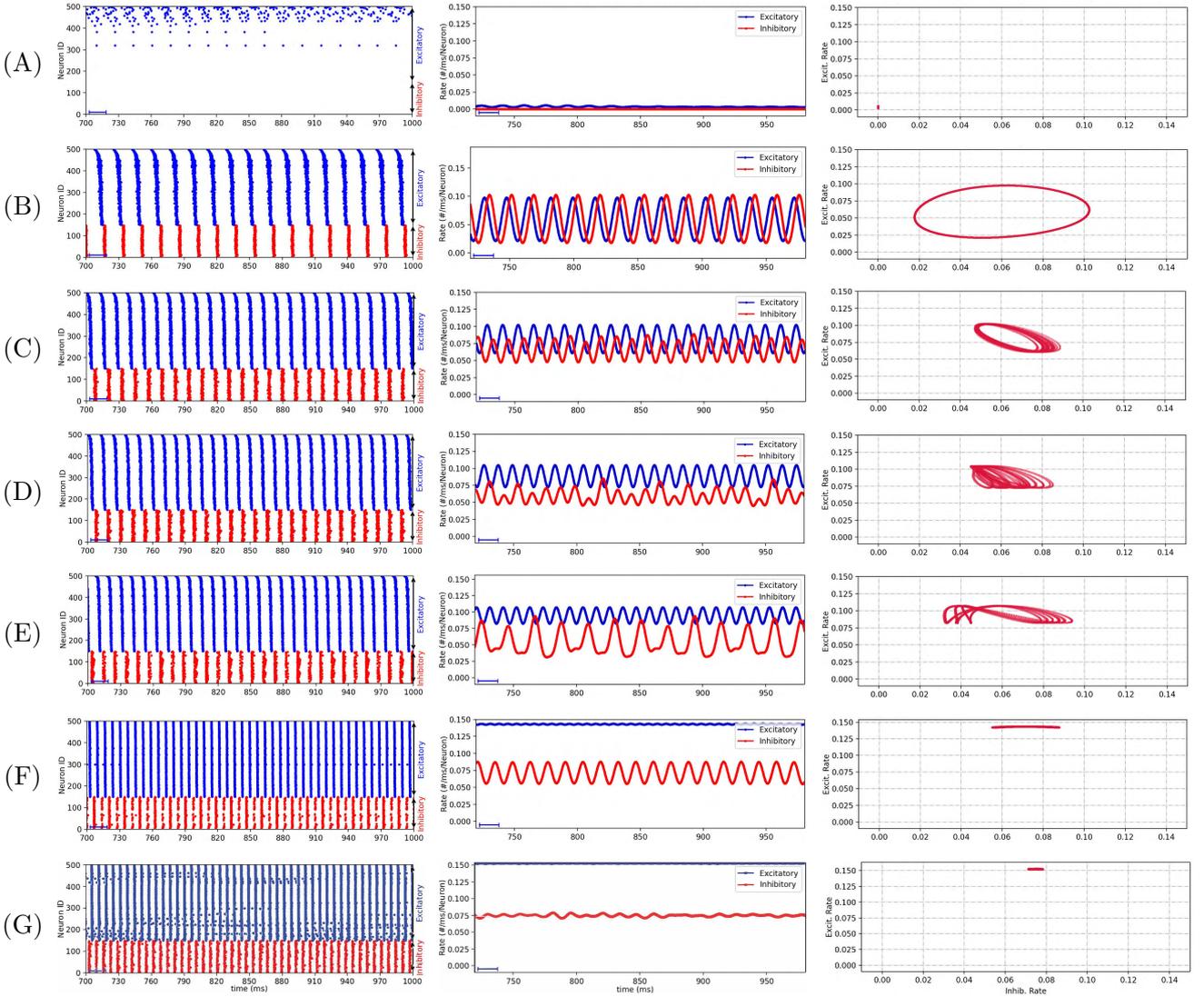

\hypertarget{Fig3}{}
\begin{center}
\begin{tabular}{@{\hspace*{0mm}}c
                @{\hspace*{2mm}}c
                @{\hspace*{2mm}}c
                @{\hspace*{2mm}}c}
(A)&
\adjustimage{width=0.31\textwidth, valign=c}{3-1-1}&
\adjustimage{width=0.31\textwidth, valign=c}{3-1-2}&
\adjustimage{width=0.31\textwidth, valign=c}{3-1-3}
\medskip
\\
(B)&
\adjustimage{width=0.31\textwidth, valign=c}{3-2-1}&
\adjustimage{width=0.31\textwidth, valign=c}{3-2-2}&
\adjustimage{width=0.31\textwidth, valign=c}{3-2-3}
\medskip
\\
(C)&
\adjustimage{width=0.31\textwidth, valign=c}{3-3-1}&
\adjustimage{width=0.31\textwidth, valign=c}{3-3-2}&
\adjustimage{width=0.31\textwidth, valign=c}{3-3-3}
\medskip
\\
(D)&
\adjustimage{width=0.31\textwidth, valign=c}{3-4-1}&
\adjustimage{width=0.31\textwidth, valign=c}{3-4-2}&
\adjustimage{width=0.31\textwidth, valign=c}{3-4-3}
\medskip
\\
(E)&
\adjustimage{width=0.31\textwidth, valign=c}{3-5-1}&
\adjustimage{width=0.31\textwidth, valign=c}{3-5-2}&
\adjustimage{width=0.31\textwidth, valign=c}{3-5-3}
\medskip
\\
(F)&
\adjustimage{width=0.31\textwidth, valign=c}{3-6-1}&
\adjustimage{width=0.31\textwidth, valign=c}{3-6-2}&
\adjustimage{width=0.31\textwidth, valign=c}{3-6-3}
\medskip
\\
(G)&
\adjustimage{width=0.31\textwidth, valign=c}{3-7-1}&
\adjustimage{width=0.31\textwidth, valign=c}{3-7-2}&
\adjustimage{width=0.31\textwidth, valign=c}{3-7-3}
\end{tabular}
\end{center}
\caption{\footnotesize\textbf{Dynamical response of the oscillatory micro population in a wide range of the input mean parameter $\bar{I}_{\textrm{stim}}$.} First column: Raster plots of the activity for $500$ Hodgkin-Huxley neurons sorted by input intensity. Second column: Evolution of state variables $Rate (t)$ of the EI system. Third column: Phase planes showing temporal trajectories in response to different values of constant excitatory current. Besides the binary re-switch between two distinct dynamical states, our neural network exhibit quasi-periodic and aperiodic orbits, as well as simple periodic solutions provided by the original Wilson-Cowan equations. The specific parameter defining the input mean is {\bf (A)} $\bar{I}_{\textrm{stim}} = 1$ \si{\micro\ampere/\square{\centi\meter}}, {\bf (B)} $\bar{I}_{\textrm{stim}} = 5$ \si{\micro\ampere/\square{\centi\meter}}, {\bf (C)} $\bar{I}_{\textrm{stim}} = 15$ \si{\micro\ampere/\square{\centi\meter}}, {\bf (D)} $\bar{I}_{\textrm{stim}} = 20$ \si{\micro\ampere/\square{\centi\meter}}, {\bf (E)} $\bar{I}_{\textrm{stim}} = 25$ \si{\micro\ampere/\square{\centi\meter}}, {\bf (F)} $\bar{I}_{\textrm{stim}} = 80$ \si{\micro\ampere/\square{\centi\meter}}, and {\bf (G)} $\bar{I}_{\textrm{stim}} = 95$ \si{\micro\ampere/\square{\centi\meter}}, and other parameter values are the input variance $\sigma_{I_{\textrm{stim}}}^{2} = 10$ \si{\micro\ampere/\square{\centi\meter}} and the global synaptic weight $W = 0.012$ \si{\siemens/\square{\centi\meter}}.}
\end{figure}
\begin{figure}[!t]
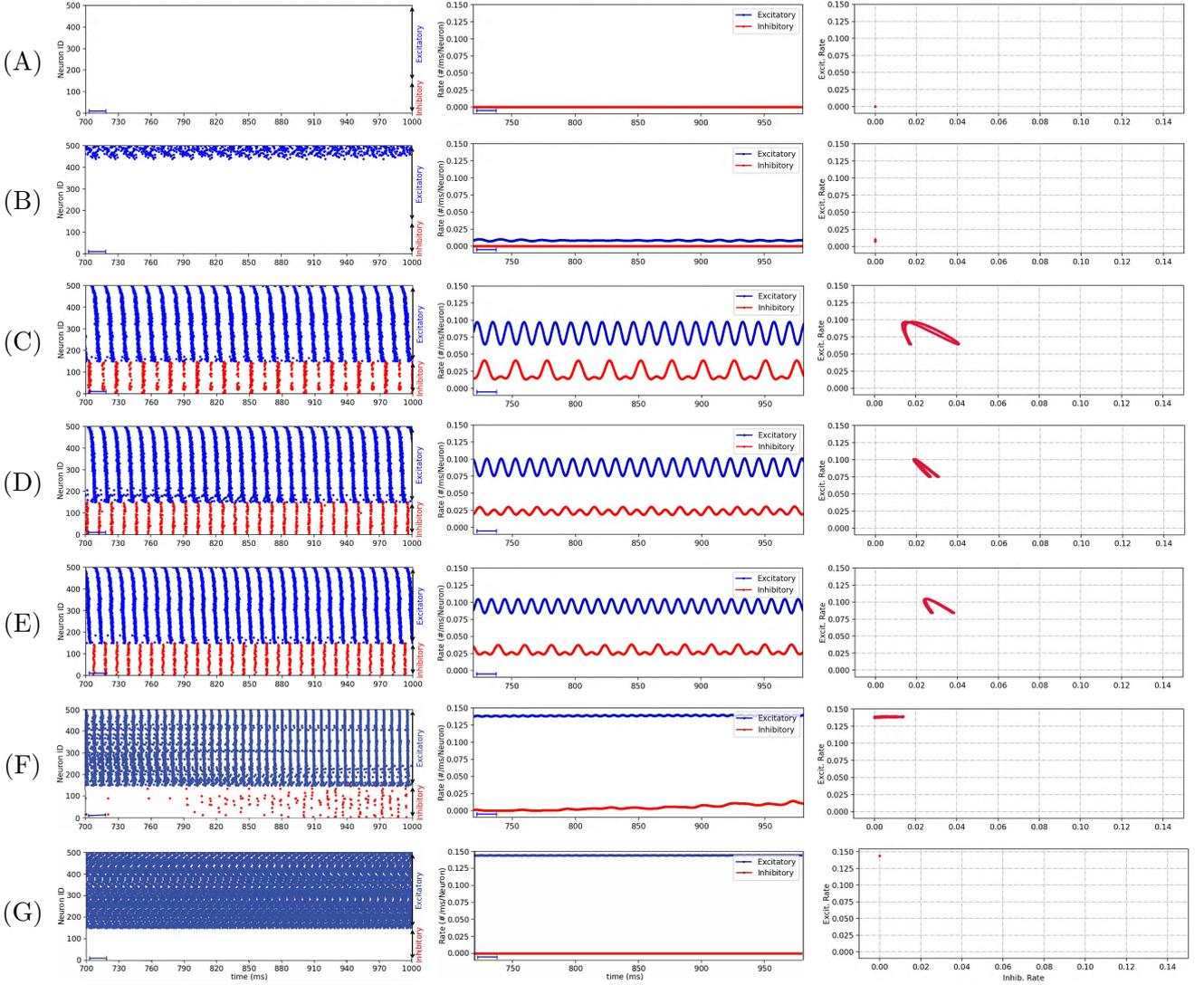

\hypertarget{Fig4}{}
\begin{center}
\begin{tabular}{@{\hspace*{0mm}}c
                @{\hspace*{2mm}}c
                @{\hspace*{2mm}}c
                @{\hspace*{2mm}}c}
(A)&
\adjustimage{width=0.31\textwidth,valign=c}{4-1-1}&
\adjustimage{width=0.31\textwidth,valign=c}{4-1-2}&
\adjustimage{width=0.31\textwidth,valign=c}{4-1-3}
\medskip
\\
(B)&
\adjustimage{width=0.31\textwidth,valign=c}{4-2-1}&
\adjustimage{width=0.31\textwidth,valign=c}{4-2-2}&
\adjustimage{width=0.31\textwidth,valign=c}{4-2-3}
\medskip
\\
(C)&
\adjustimage{width=0.31\textwidth,valign=c}{4-3-1}&
\adjustimage{width=0.31\textwidth,valign=c}{4-3-2}&
\adjustimage{width=0.31\textwidth,valign=c}{4-3-3}
\medskip
\\
(D)&
\adjustimage{width=0.31\textwidth,valign=c}{4-4-1}&
\adjustimage{width=0.31\textwidth,valign=c}{4-4-2}&
\adjustimage{width=0.31\textwidth,valign=c}{4-4-3}
\medskip
\\
(E)&
\adjustimage{width=0.31\textwidth,valign=c}{4-5-1}&
\adjustimage{width=0.31\textwidth,valign=c}{4-5-2}&
\adjustimage{width=0.31\textwidth,valign=c}{4-5-3}
\medskip
\\
(F)&
\adjustimage{width=0.31\textwidth,valign=c}{4-6-1}&
\adjustimage{width=0.31\textwidth,valign=c}{4-6-2}&
\adjustimage{width=0.31\textwidth,valign=c}{4-6-3}
\medskip
\\
(G)&
\adjustimage{width=0.31\textwidth,valign=c}{4-7-1}&
\adjustimage{width=0.31\textwidth,valign=c}{4-7-2}&
\adjustimage{width=0.31\textwidth,valign=c}{4-7-3}
\end{tabular}
\end{center}
\caption{\footnotesize Simulation results of the EI network with same parameters as in Figure\,\protect\hyperlink{Fig3}{3} except for $W = 0.005$ \si{\siemens/\square{\centi\meter}}. The low coupling strength causes a lower degree of synchrony, and tunes the neural system near the different types of fixed points.}
\end{figure}
\begin{figure}[!t]
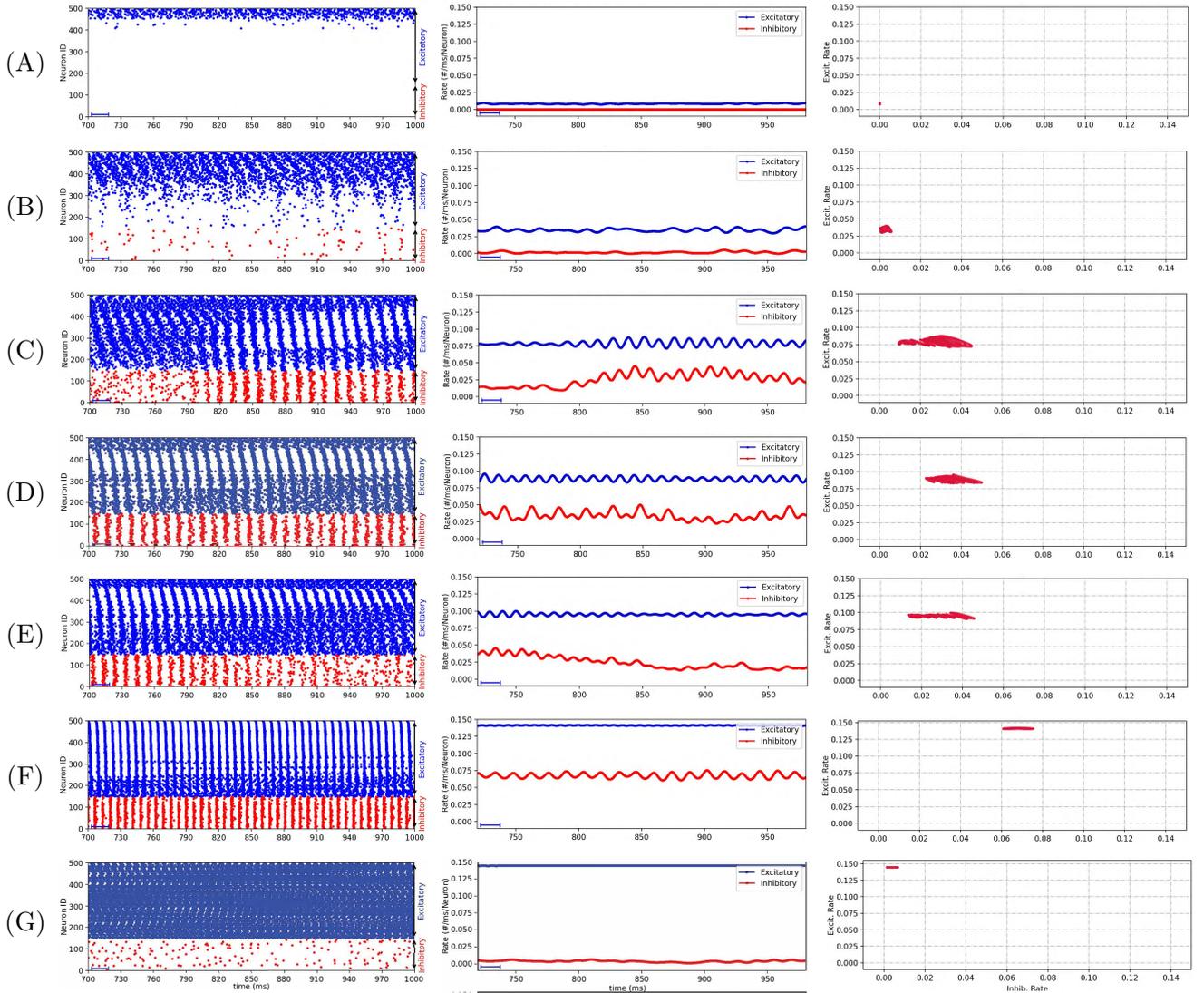

\hypertarget{Fig5}{}
\begin{center}
\begin{tabular}{@{\hspace*{0mm}}c
                @{\hspace*{2mm}}c
                @{\hspace*{2mm}}c
                @{\hspace*{2mm}}c}
(A)&
\adjustimage{width=0.31\textwidth,valign=c}{5-1-1}&
\adjustimage{width=0.31\textwidth,valign=c}{5-1-2}&
\adjustimage{width=0.31\textwidth,valign=c}{5-1-3}
\medskip
\\
(B)&
\adjustimage{width=0.31\textwidth,valign=c}{5-2-1}&
\adjustimage{width=0.31\textwidth,valign=c}{5-2-2}&
\adjustimage{width=0.31\textwidth,valign=c}{5-2-3}
\medskip
\\
(C)&
\adjustimage{width=0.31\textwidth,valign=c}{5-3-1}&
\adjustimage{width=0.31\textwidth,valign=c}{5-3-2}&
\adjustimage{width=0.31\textwidth,valign=c}{5-3-3}
\medskip
\\
(D)&
\adjustimage{width=0.31\textwidth,valign=c}{5-4-1}&
\adjustimage{width=0.31\textwidth,valign=c}{5-4-2}&
\adjustimage{width=0.31\textwidth,valign=c}{5-4-3}
\medskip
\\
(E)&
\adjustimage{width=0.31\textwidth,valign=c}{5-5-1}&
\adjustimage{width=0.31\textwidth,valign=c}{5-5-2}&
\adjustimage{width=0.31\textwidth,valign=c}{5-5-3}
\medskip
\\
(F)&
\adjustimage{width=0.31\textwidth,valign=c}{5-6-1}&
\adjustimage{width=0.31\textwidth,valign=c}{5-6-2}&
\adjustimage{width=0.31\textwidth,valign=c}{5-6-3}
\medskip
\\
(G)&
\adjustimage{width=0.31\textwidth,valign=c}{5-7-1}&
\adjustimage{width=0.31\textwidth,valign=c}{5-7-2}&
\adjustimage{width=0.31\textwidth,valign=c}{5-7-3}
\end{tabular}
\end{center}
\caption{\footnotesize Simulation results of the EI network with same parameters as in Figure\,\protect\hyperlink{Fig3}{3} except that
here $\sigma_{I_{\textrm{stim}}}^{2} = 40$ \si{\micro\ampere/\square{\centi\meter}}. The high variance in the input distribution prevents the enhancement of local synchrony, and tunes the neural system near the low and high rate fixed points.}
\end{figure}
\subsection{The interplay of collective behaviour and excitatory structural organization on the microscale level}
\hypertarget{Res3}{}

We analyze and simulate the behavioural dynamics of the spiking neural network and its bifurcations under the different values of major topological parameters, i.e., the local network degrees quantifying the attractive and repulsive interconnections between excitatory and inhibitory subpopulations.

The average degree of E-E connectivity is an important contributor to generate periodic patterns in the locally coupled oscillators undergoing a constant stimulus, that is, the range of the excitatory type coupling is encompassed within a special domain and, subsequently, limit cycle dynamics might not be accessible to the system outside this scenario (Figure\,\hyperlink{Fig6}{6\,A}). This is reasonable, since sufficient interactions within the excitatory population tend to regulate the strength of inhibitory feedback loop which is responsible for the occurrence of a Hopf bifurcation at a quiescent fixed point. As given in Figure\,\hyperlink{Fig6}{6\,B}, increasing the self-excitation E-E connections can push the microscopic network of neurons over the level of their firing threshold, and favors in-phase synchronization and clustering reflected by the spike record. It can also be speculated that the larger the number of excitatory synapses between excitatory neurons, the higher the amplitude of population firing rates and the corresponding limit cycle oscillations, as well as the lower the frequency of these dynamical patterns (Figures\,\hyperlink{Fig6}{6\,B}, \hyperlink{Fig6}{C}, \hyperlink{Fig6}{D}, and \hyperlink{Fig6}{E}). Considering the synaptically coupled system in a regime with few E-I synapses and under a proper external force, this combination places the system on a path of oscillatory dynamics which is possible as one of the outcomes in the simplest setting of the Wilson-Cowan populations (Figures\,\hyperlink{Fig7}{7\,A}, \hyperlink{Fig7}{B}, and \hyperlink{Fig7}{C}). This is so because the inhibitory subpopulation is excited only by the excitatory subpopulation, thus we would expect the E-I connections among neurons to have considerable influence on governing the complexity of the underlying neural dynamics and shaping the temporal behaviour of localized EI population. Notably, according to Figures\,\hyperlink{Fig7}{7\,A}-\hyperlink{Fig7}{E}, all neurons in the excitatory pool operate in a synchronous regular regimen with oscillatory spiking rate but without attending to dependence on the E-I coupling's degree, although the coordination of inhibitory oscillators is positively correlated to this type of connectomes. And, owing to the self-excitatory loop and the related E-I excitation, the connectivity architecture of inhibitory synapses, plus the negative loop, does not play a significant role for generating and maintaining the coherent activity of our neural ensemble (see Figure\,S1 and Figure\,S2).

These findings suggest that the coupled  microscopic system shows an apparent tendency towards the macroscopic mean-field treatment when increasing the local excitatory connections distributed uniformly between the individual neurons. In other words, if heterogeneity in the excitatory synaptic configuration is minimized, the Wilson-Cowan oscillations, which arise from a Hopf bifurcation, can emerge at the large-scale network level. In line with the previous results, also here the population of binary threshold oscillators relying on the attractive coupling is able to drive an up-down phase transition state which does not follow the predictions of the mean-field deterministic trajectory.
\begin{figure}[!t]
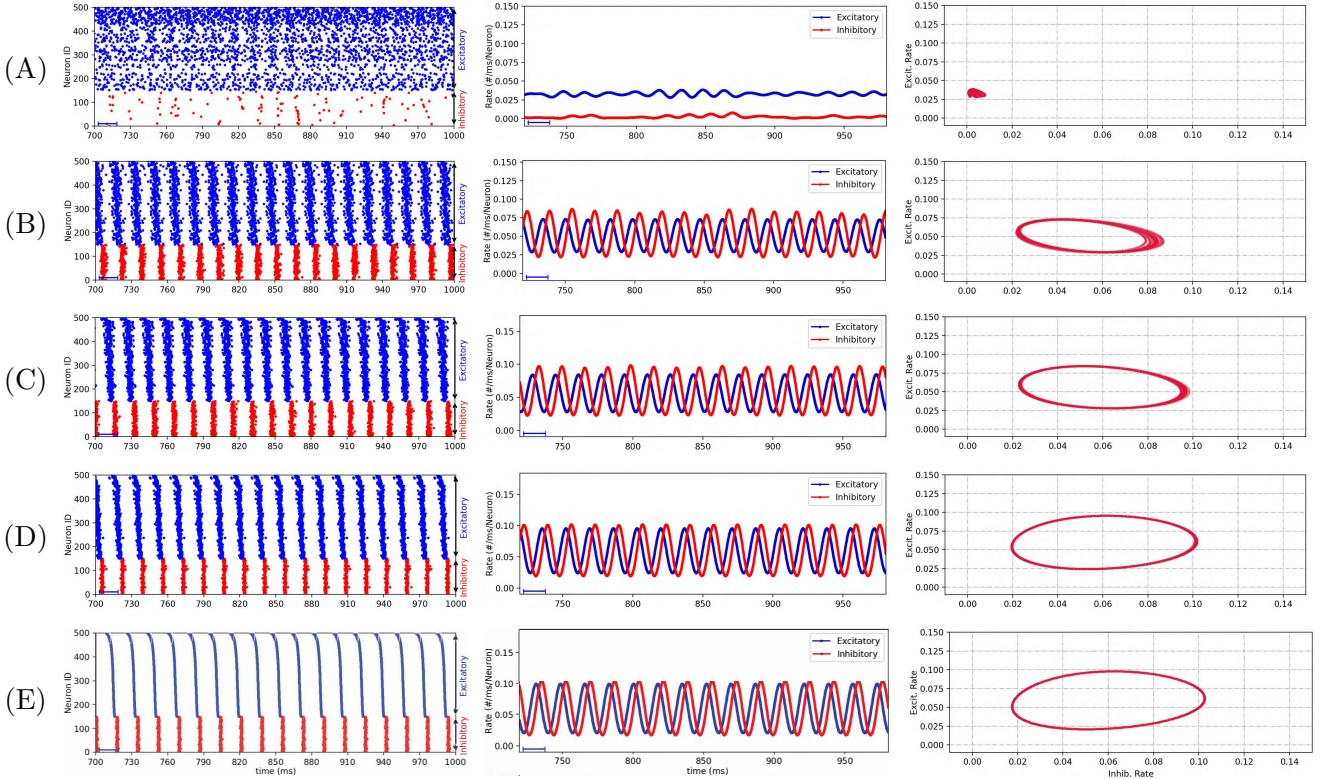

\hypertarget{Fig6}{}
\begin{center}
\begin{tabular}{@{\hspace*{0mm}}c
                @{\hspace*{2mm}}c
                @{\hspace*{2mm}}c
                @{\hspace*{2mm}}c}
(A)&
\adjustimage{width=0.31\textwidth,valign=c}{6-1-1}&
\adjustimage{width=0.31\textwidth,valign=c}{6-1-2}&
\adjustimage{width=0.31\textwidth,valign=c}{6-1-3}
\medskip
\\
(B)&
\adjustimage{width=0.31\textwidth,valign=c}{6-2-1}&
\adjustimage{width=0.31\textwidth,valign=c}{6-2-2}&
\adjustimage{width=0.31\textwidth,valign=c}{6-2-3}
\medskip
\\
(C)&
\adjustimage{width=0.31\textwidth,valign=c}{6-3-1}&
\adjustimage{width=0.31\textwidth,valign=c}{6-3-2}&
\adjustimage{width=0.31\textwidth,valign=c}{6-3-3}
\medskip
\\
(D)&
\adjustimage{width=0.31\textwidth,valign=c}{6-4-1}&
\adjustimage{width=0.31\textwidth,valign=c}{6-4-2}&
\adjustimage{width=0.31\textwidth,valign=c}{6-4-3}
\medskip
\\
(E)&
\adjustimage{width=0.31\textwidth,valign=c}{6-5-1}&
\adjustimage{width=0.31\textwidth,valign=c}{6-5-2}&
\adjustimage{width=0.31\textwidth,valign=c}{6-5-3}
\end{tabular}
\end{center}
\caption{\footnotesize\textbf{Qualitative relationship between degree of excitatory type coupling and collective dynamics of the microscale neurons.} First column: Raster plots of the activity for $500$ Hodgkin-Huxley neurons sorted by input intensity. Second column: Evolution of state variables $Rate (t)$ of the EI system. Third column: Phase planes showing temporal trajectories in response to constant stimulation. The local excitatory connectivity E-E can help the localized neuronal population to achieve all-to-all network's behaviour comparable to that of the traditional Wilson-Cowan oscillators. The specific parameter defining the mean of E-E coupling is {\bf (A)} $\mu_{\textrm{E-E}} = 1$, {\bf (B)} $\mu_{\textrm{E-E}} = 4$, {\bf (C)} $\mu_{\textrm{E-E}} = 6$, {\bf (D)} $\mu_{\textrm{E-E}} = 10$, and {\bf (E)} $\mu_{\textrm{E-E}} = 350$, and other parameter value is the variance of E-E coupling $\sigma_{\textrm{E-E}}^{2} = 3$.}
\end{figure}
\begin{figure}[!t]
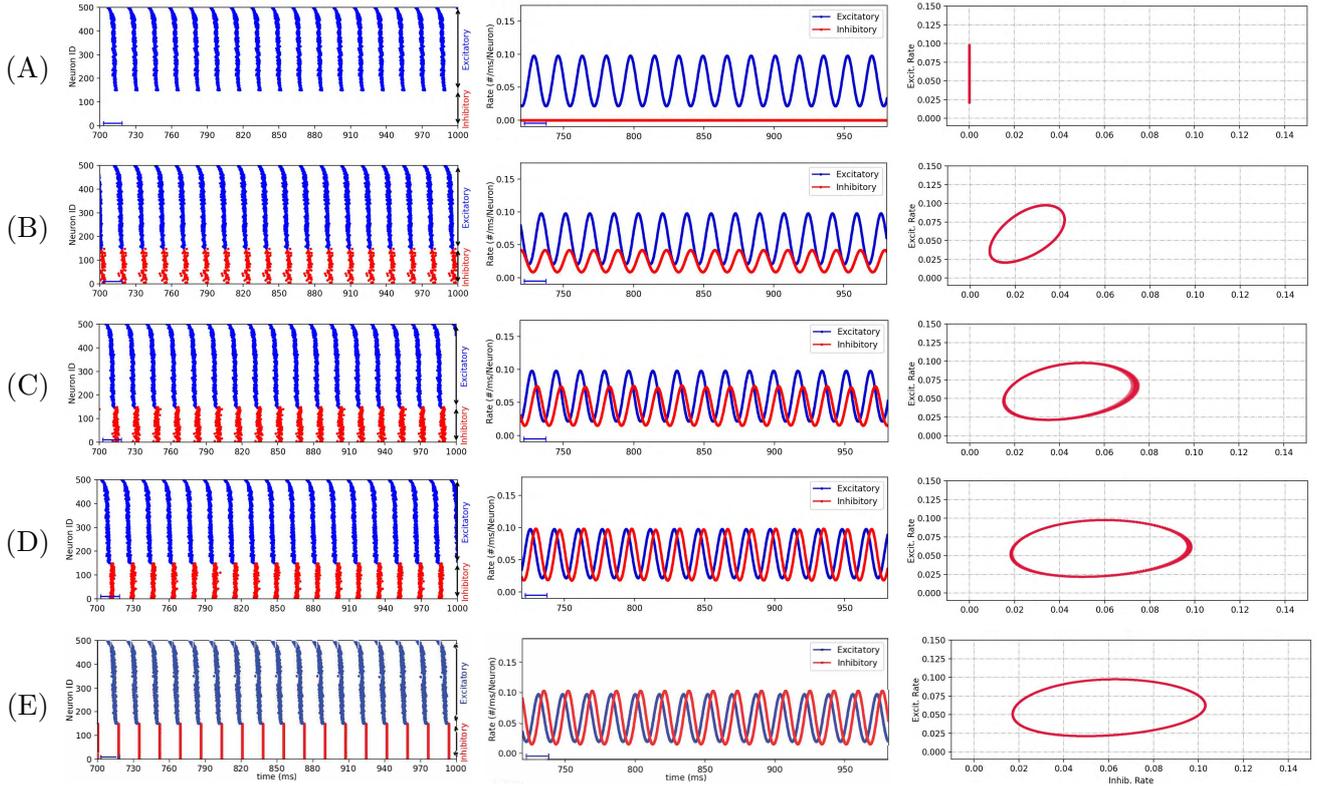

\hypertarget{Fig7}{}
\begin{center}
\begin{tabular}{@{\hspace*{0mm}}c
                @{\hspace*{2mm}}c
                @{\hspace*{2mm}}c
                @{\hspace*{2mm}}c}
(A)&
\adjustimage{width=0.31\textwidth,valign=c}{7-1-1}&
\adjustimage{width=0.31\textwidth,valign=c}{7-1-2}&
\adjustimage{width=0.31\textwidth,valign=c}{7-1-3}
\medskip
\\
(B)&
\adjustimage{width=0.31\textwidth,valign=c}{7-2-1}&
\adjustimage{width=0.31\textwidth,valign=c}{7-2-2}&
\adjustimage{width=0.31\textwidth,valign=c}{7-2-3}
\medskip
\\
(C)&
\adjustimage{width=0.31\textwidth,valign=c}{7-3-1}&
\adjustimage{width=0.31\textwidth,valign=c}{7-3-2}&
\adjustimage{width=0.31\textwidth,valign=c}{7-3-3}
\medskip
\\
(D)&
\adjustimage{width=0.31\textwidth,valign=c}{7-4-1}&
\adjustimage{width=0.31\textwidth,valign=c}{7-4-2}&
\adjustimage{width=0.31\textwidth,valign=c}{7-4-3}
\medskip
\\
(E)&
\adjustimage{width=0.31\textwidth,valign=c}{7-5-1}&
\adjustimage{width=0.31\textwidth,valign=c}{7-5-2}&
\adjustimage{width=0.31\textwidth,valign=c}{7-5-3}
\end{tabular}
\end{center}
\caption{\footnotesize Simulation results of the EI network with same parameters as in Figure\,\protect\hyperlink{Fig6}{6} except for the mean of E-I coupling {\bf (A)} $\mu_{\textrm{E-I}} = 1$, {\bf (B)} $\mu_{\textrm{E-I}} = 4$, {\bf (C)} $\mu_{\textrm{E-I}} = 6$, {\bf (D)} $\mu_{\textrm{E-I}} = 10$, and {\bf (E)} $\mu_{\textrm{E-I}} = 350$, and the variance of E-I coupling $\sigma_{\textrm{E-I}}^{2} = 3$. The local excitatory connectivity E-I can help the localized neuronal population to achieve all-to-all network's behaviour comparable to that of the traditional Wilson-Cowan oscillators.}
\end{figure}
\subsection{The effects of topology and external current in the dynamical repertoire of neural behaviours}
\hypertarget{Res4}{}

We show phase portraits of the present conductance-based model in the $(W, \bar{I}_{\textrm{stim}})$-parameter space, which correspond to different interconnection structures ranging from topologically invariant random lattices to random networks with changing the structural geometry and using the different classes of stimuli (cf. Figure\,\hyperlink{Fig8}{8}-Figure\,\hyperlink{Fig10}{10}; see also Figure\,S3). The potential advantage to be gained from the schematic representations is the applicability of mathematical analysis to extract remarkable qualitative features of the EI population's mean activity for various parameter ranges.

First, it is important to notice that all the considered topologies are capable of performing a rich repertoire of dynamical modes including the different types of fixed points, limit cycles or sustained oscillations, quasi-periodic orbits or damped oscillations, chaotic dynamics, and sharp phase transitions from silent to active states; however, the last three are not substantially born from the Wilson-Cowan equations in their most simple incarnation. Now, assuming the degree distribution of local population to be varied per point $(W, \bar{I}_{\textrm{stim}})$ in the control parameter plane, a sketch of the phase space from different regimes is depicted in Figure\,\hyperlink{Fig8}{8}. Compared to the dynamical portrait of topology-invariant version of the network, given Figure\,S3, the alterations in the temporal profile remain nearly similar in both these cases. This implies that the system is less sensitive to the degree of randomness in the connectivity because of the Gaussian approximation of the synaptic distribution. A decrease in the variance of external drive makes more rows of the phase scheme susceptible to capture the emergence of stable quiescent fixed points due to insufficient excitation (Figure\,\hyperlink{Fig9}{9} vs. Figure\,\hyperlink{Fig8}{8}). Nevertheless, as illustrated in Figure\,\hyperlink{Fig10}{10}, increasing the variance of excitatory input raises the firing noise probability to exceed some threshold value which leads to the earlier appearance of limit cycle behaviour in low stimulus levels. This emphasizes the importance of having considerable diversity in the external current in order to locate the Hopf bifurcation points, as well as trigger stable local oscillations and synchrony in the microscopic network. Moreover, following a similar $(W, \bar{I}_{\textrm{stim}})$-dependent pattern in all the phase repertoires, the activity correlation of excitatory and inhibitory oscillators is firstly positive and then tend to change toward negative, thereby affecting the direction of phase difference between them.
\begin{figure}[!t]
\hypertarget{Fig8}{}
\centering
{
\includegraphics[width=\textwidth]{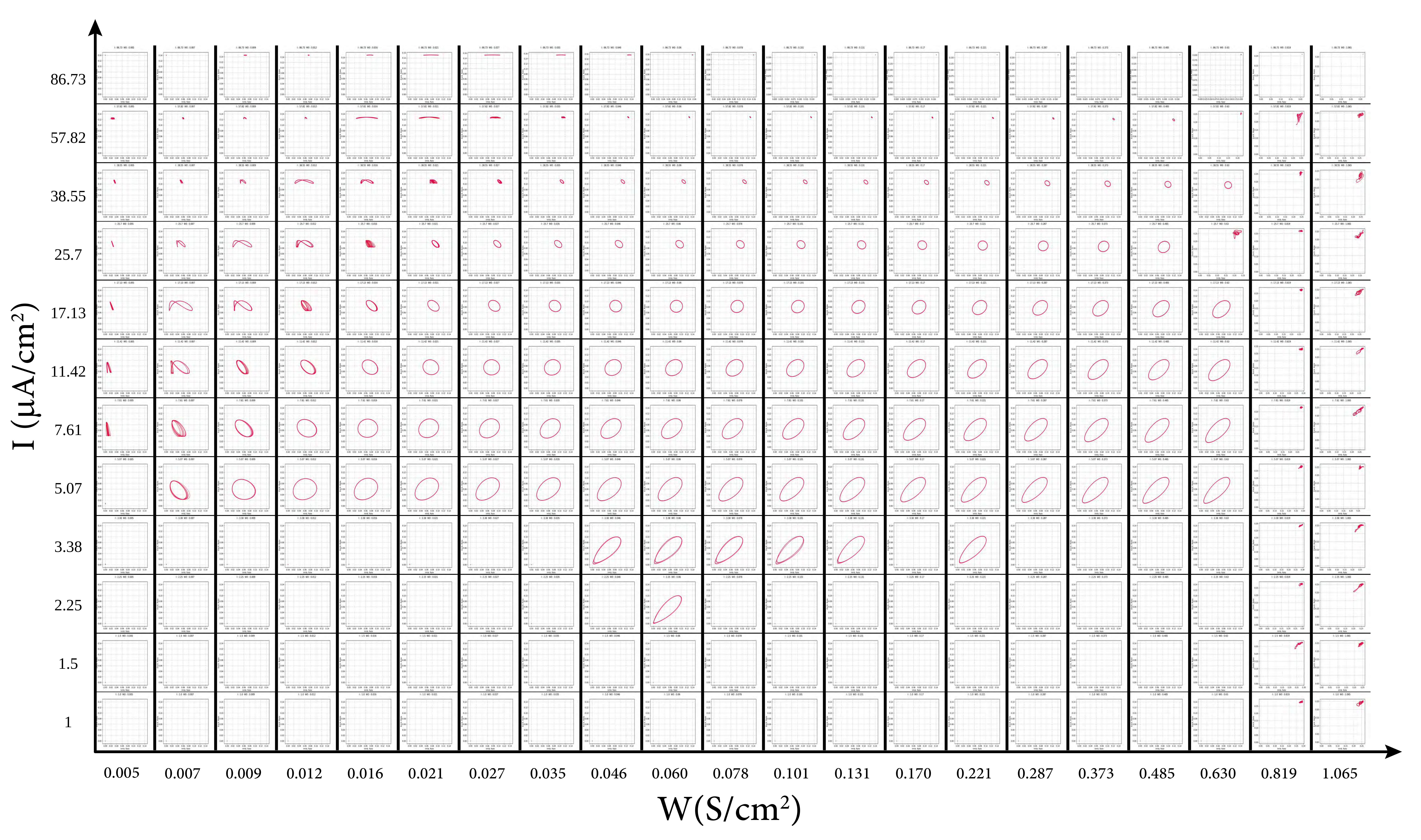}
}
\vspace{-7mm}
\caption{\footnotesize\textbf{$(W, \bar{I}_{\textrm{stim}})$-phase portrait of topology-variant version of the random EI networks in log-log plot.} The schematic representation completely covers different temporal dynamics ranging from simple periodic to quasi-periodic and aperiodic patterns, as well as phase transition regimes. The specific parameter defining the input variance is $\sigma_{I_{\textrm{stim}}}^{2} = 10$ \si{\micro\ampere/\square{\centi\meter}}.}
\end{figure}
\begin{figure}[!t]
\hypertarget{Fig9}{}
\centering
{
\includegraphics[width=\textwidth]{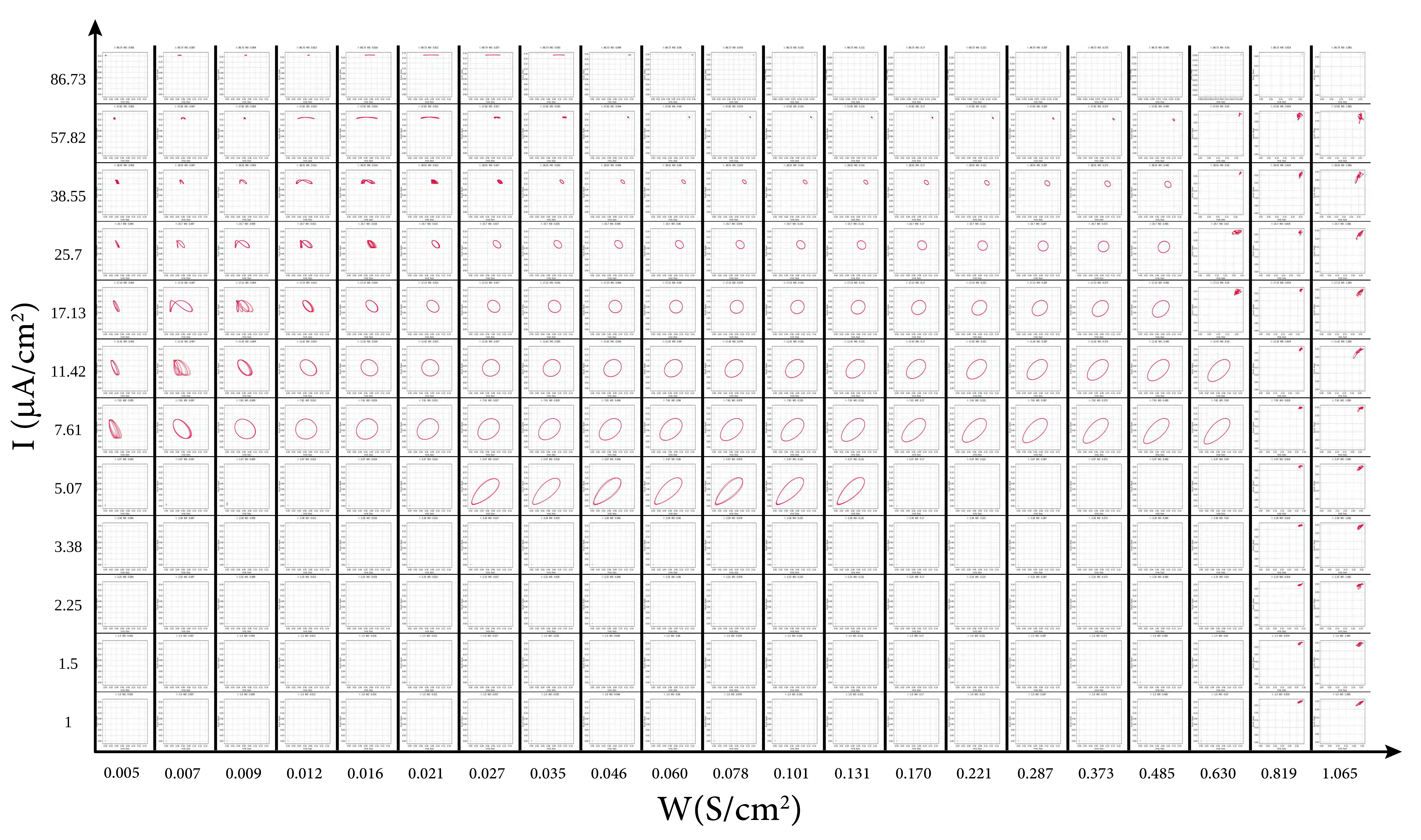}
}
\vspace{-7mm}
\caption{\footnotesize Same as Figure\,\protect\hyperlink{Fig8}{8}, illustrating $(W, \bar{I}_{\textrm{stim}})$-phase portrait of topology-variant version of the random EI networks in log-log plot, but with a new value for the input variance $\sigma_{I_{\textrm{stim}}}^{2} = 5$ \si{\micro\ampere/\square{\centi\meter}}. Decresing the variance of excitatory input decreases the firing probability to exceed some threshold value which leads to the later appearance of limit cycle behaviour in higher stimulus levels.}
\end{figure}
\begin{figure}[!t]
\hypertarget{Fig10}{}
\centering
{
\includegraphics[width=\textwidth]{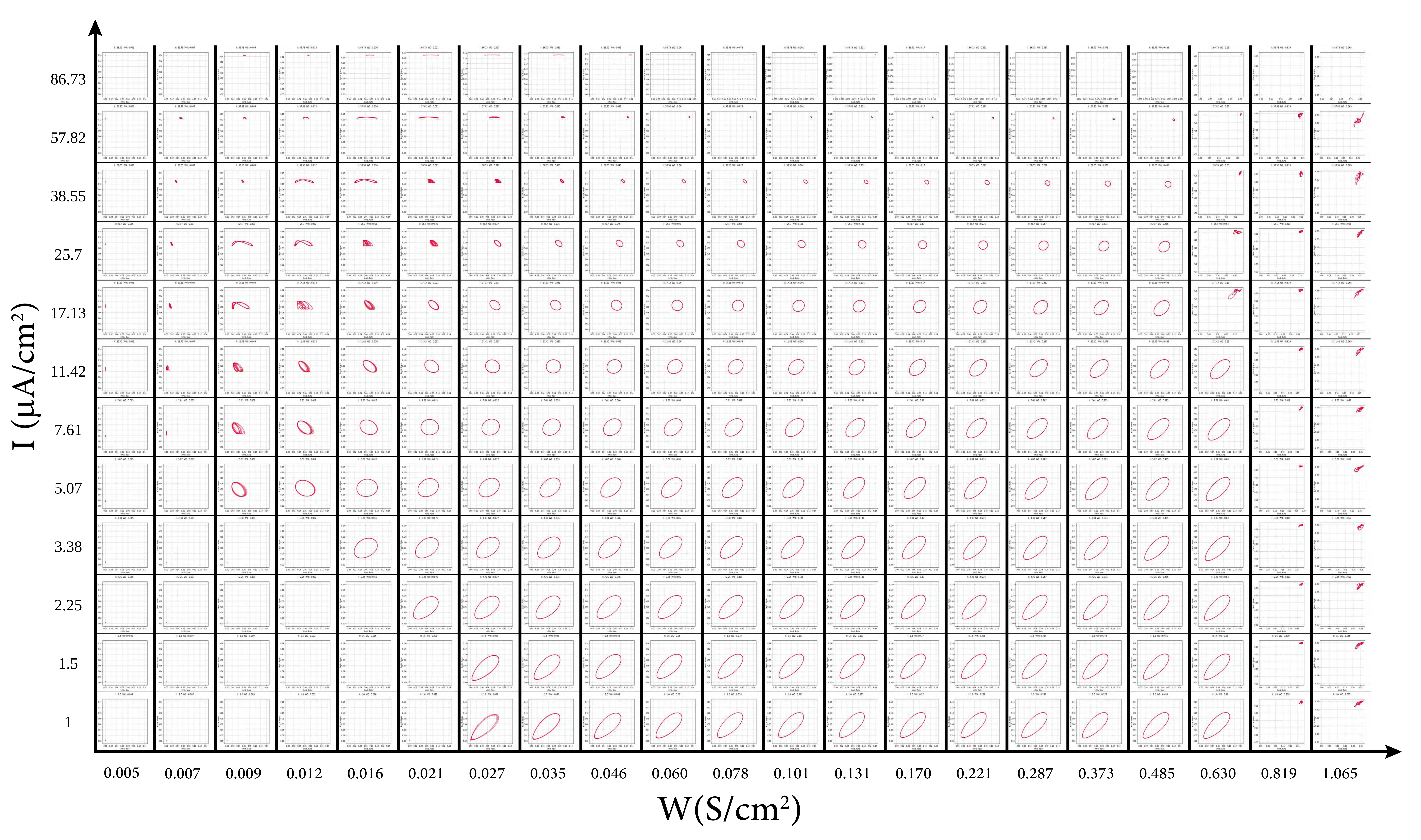}
}
\vspace{-7mm}
\caption{\footnotesize Same as Figure\,\protect\hyperlink{Fig8}{8}, illustrating $(W, \bar{I}_{\textrm{stim}})$-phase portrait of topology-variant version of the random EI networks in log-log plot, but with a new value for the input variance $\sigma_{I_{\textrm{stim}}}^{2} = 20$ \si{\micro\ampere/\square{\centi\meter}}. An increase in the variance of external drive makes more rows of the phase scheme susceptible to capture the emergence of limit cycle trajectories due to sufficient excitation.}
\end{figure}
\section{DISCUSSION}
\hypertarget{Dis}{}

In the present work, we develop a unified nonlinear-dynamical network model to explore numerically whether or not the traditional Wilson-Cowan equations, as a firing-rate mass model, can successfully tackle a comprehensive dynamic landscape at the neural population level. The model combines biophysical constraints from a neural-level model (Hodgkin-Huxley) with additional mean-field assumptions from a population-level model (Wilson-Cowan) regarding structural homogeneities, internal interactions, as well as stimulus configurations. While an increase in the synaptic strength always enhances the degree of synchronization \cite{Bennett(2004),Buzsaki(2006)}, compared to the macroscopic mass dynamics in terms of the coupling parameter, the temporal profile in the microscopic system raises substantial discrepancies manifested predominantly in the Hopf bifurcations at both the low and high rates (Figure\,\hyperlink{Fig2}{2}). Interestingly, ensuring a rather consistency with the simplest Wilson-Cowan reference, the synaptic weight close to these down and up bifurcations drives the micro EI population to different dynamical regimes in which the system wanders around the active-inactive phase transitions. This phenomenon might be attributed to the binary state of the Hodgkin-Huxley model and the random distribution for the connection topology, which are not explicitly included in the Wilson-Cowan trajectory. Although Kaslik and co-workers did exhibit a generalization of the Wilson-Cowan type system in which the transition regimes were governed by both the coupling strength and the shape of delay distribution \cite{Kaslik(2021),Kaslik(2020)}, we want to emphasize that the more general outcome of a neural population via its mass dynamics and the mean-field dependence may seem less realistic and thus requires great care. Further studies support this claim in different setups like the Freeman neural-mass model with a network of leaky integrate-and-fire neurons \cite{Deschle(2021)} and the macroscopic Wilson-Cowan dynamics with biophysical constraints of the Wong-Wang-Deco model \cite{Zhang(2020)}. On the other hand, phase transitions and bifurcations have attracted a lot of interest since they seem to particularly characterize the complexity of resting state dynamics \cite{Ito(2005),Ito(2007)}, may describe ubiquitous multistability and metastability at whole-brain level \cite{Tognoli(2014),Deco(2017),Beim Graben(2019)}, and also be linked to generalized seizure states \cite{Breakspear(2006),Rodrigues(2009),Marten(2009)}.

In the general case, the Wilson-Cowan model of mean-field EI interactions can easily generate limit cycle dynamics in response to appropriate constant stimulation \cite{Wilson(1972),Wilson(1973)}. In our microscopic network model, in addition to the binary re-switch between the input-driven dynamics of qualitatively distinct states, we observe that there exist the quasi-periodic and aperiodic orbits in the $(\textrm{Excit. Rate}, \textrm{Inhib. Rate})$-phase plane for the reasonable values of input rate which, despite imposing the mean-field restrictions, not correspond fundamentally to the original Wilson-Cowan's oscillations (Figure\,\hyperlink{Fig3}{3}). It should be noted that Borisyuk et al. \cite{Borisyuk(1995)} and Maruyama et al. \cite{Maruyama(2014)} extended the Wilson-Cowan equations to include a pair of coupled Wilson-Cowan oscillators which lead to produce chaos trajectory, especially if inhibitory input is from each external oscillator to excitatory interneurons of the adjacent oscillator. And Wilson \cite{Wilson(2019)} and Kaslik et al. \cite{Kaslik(2021),Kaslik(2020)} respectively demonstrated that chains of Wilson-Cowan oscillators with inhibitory coupling are effective at generating hyperchaos regions of quite high level, and the Wilson-Cowan with distributed time delays can reveal complex bursting and chaotic behaviour. But, this result points out the simplified setting of such neural-mass models may significantly alter the dynamical behaviour of self-interacting EI populations. In fact, as previously mentioned, we construct a simple model of the oscillatory micro population and attempt to close its general conditions to the Wilson-Cowan approximations, although exhibiting the complex temporal patterns which by eliminating some of the simplifications, more complicated scenarios can occur. Furthermore, our simulation results show, on one hand, that greatly increasing the variance of excitatory input moves the microscopic system towards asynchronous irregular regions due to the enormous diversity in the stimulation range (Figure\,\hyperlink{Fig5}{5} vs. Figure\,\hyperlink{Fig3}{3}), and on the other, that keeping the average external force extremely strong leads the firing activity to saturate due to the refractoriness
(Figures\,\hyperlink{Fig3}{3\,F} and \hyperlink{Fig3}{G}). And, the low variance or mean in the input distribution prevents the enhancement of local synchrony, and tunes the system near the fixed point quiescent states (not shown). Actually, we can conclude from the above analysis that the proper diversity of average external current to the excitatory population, together with an adequate amount of the synaptic weight, is a key contributor to the periodicity of output firing rate and the limit cycle dynamics.

Nonlinearity in the local dynamics and oscillatory in the local population's activity, consisting of damped and sustained oscillations, may be effectively controlled by the strength of local excitatory-to-excitatory connection and local excitatory-to-inhibitory connection \cite{Zhang(2020)}, as also discussed in \cite{Demirtas(2019)}. We identify such crucial role played by the average degree of local excitatory connectivity E-E and E-I in generating and amplifying Hopf bifurcation dynamics (Figure\,\hyperlink{Fig6}{6} and Figure\,\hyperlink{Fig7}{7}). This ability for such attractive interconnections, opposite to the repulsive characteristic of inhibitory synapses, can come obviously to help the system achieve all-to-all network's behaviour comparable to that of the traditional Wilson-Cowan oscillators, which is formally identified by the limit cycle regimes in the phase plane, and also to help the switching dynamics and transitions between them. The importance of inhibitory interactions both within and between localized populations in the generation of rhythmic activity and chaos plus hyperchaos regions derives from both theoretical and empirical studies, e.g. \cite{Borisyuk(1995),Maruyama(2014),Wilson(2019),Atallah(2009),Mann(2009),Middleton(2008),Bartos(2007),Wang(1996)}.

In the field of computational neuroscience, it is always a challenge to link the complexity of neural dynamics to the underlying structure. Nonlinear-dynamical models serve as important tools for appropriately analyzing structural constraints and their dynamical consequences \cite{Deco(2011),Park(2013)}. P\'{e}rez and co-workers, based on numerical simulations of conductance-based neurons by the Hodgkin-Huxley and Connor-Stevens type models, indicated that different interconnection topologies including regular, small-world, random, scale-free and globally coupled networks exhibit a coherent response at a local level; however, at a global level, only (scale-free) networks with specific degree of randomness in the connectivity are able to obtain a coordinated firing \cite{Perez(2011)}. Zhang and Saggar used a unified mean-field model of the Wilson-Cowan dynamics and the Wong-Wang-Deco biophysical constraints, and showed that multistability and temporal diversity of the whole brain are shaped jointly by the diversity of local-scale structural connectivity and the topology of large-scale network \cite{Zhang(2020)}. Here, we demonstrate that the behavioural dynamics of the microscopic network does not depend generally on the random property of local Gaussian connectivity extracted from the macroscopic Wilson-Cowan model, except for phase transition states in which the EI system wanders into these critical regions.
\section{CONCLUSION}
\hypertarget{Concl}{}

The Wilson-Cowan type neural-mass is one of the most essential tools to study a localized EI population's mean activity. It is believed to mimic selected empirical phenomena including sustained local oscillations and patterns of inhomogeneous synchrony or asynchrony. We question its validity and accuracy as a coarse-grained approximation to capture the key dynamical features of neural populations. For a large-scale network of threshold neurons with random connectivity, we find out that corresponding collective dynamics may deviate considerably from the mean-field trajectory. Deviations occur rarely close to Hopf bifurcation points through either increasing the synaptic strength or the external drive which are the origin of quasi-periodic or chaotic patterns and wandering behaviours in the microscopic system, while the original Wilson-Cowan equations exhibit just periodic solutions at the macroscale level. The phenomenological neural-mass models describing several dynamical modes require to include more than pure mean-field assumptions, since these modeling approaches capture the relevant mean treatment while discounting fluctuations around the mean, and do not enforce any set of the biophysical constraints on their equations.
\section{METHODS}
\hypertarget{Meth}{}
\subsection{Neuron dynamics}

To simulate the collective behaviour of $N = 500$ neurons coupled with chemical synapses, we apply a conductance-based model (Hodgkin-Huxley). The sub-threshold dynamics of the membrane potential $V^{(\textrm{i})}_{\!\textrm{m}}$ of the $\textrm{i}$-th neuron $(1 \le \textrm{i} \le N)$ is described as follows
\begin{equation*}
C_{\textrm{m}} \dfrac{dV^{(\textrm{i})}_{\!\textrm{m}}}{dt} = - I^{(\textrm{i})}_{\textrm{m}} - I^{(\textrm{i})}_{\textrm{syn}} + I^{(\textrm{i})}_{\textrm{stim}},
\end{equation*}
where $C_{\textrm{m}}$ is the membrane capacitance, $I^{(\textrm{i})}_{\textrm{m}}$ represents the membrane current, $I^{(\textrm{i})}_{\textrm{syn}}$ is the set of synaptic currents coming from neighboring neurons, and $I^{(\textrm{i})}_{\textrm{stim}}$ stands for the external stimulated current.

In the classical Hodgkin-Huxley formalism \cite{Hodgkin(1952)}, the ionic current flowing across the nerve membrane satisfies
\begin{equation*}
I^{(\textrm{i})}_{\textrm{m}} = g_{\textrm{Na}} m^{3} h ( V^{(\textrm{i})}_{\!\textrm{m}} - E_{\textrm{Na}}) + g_{\textrm{K}} n^{4} ( V^{(\textrm{i})}_{\!\textrm{m}} - E_{\textrm{K}}) + g_{\textrm{leak}} ( V^{(\textrm{i})}_{\!\textrm{m}} - E_{\textrm{leak}}),
\end{equation*}
where $g_{\textrm{p}}$ with $\textrm{p} = \{ \textrm{Na}, \textrm{K}, \textrm{leak} \}$ indicate the maximal conductance density related to the ionic contributions and passive channel, respectively, and $E_{\textrm{p}}$ give the corresponding Nernst potentials (or reversal potentials). The parameters $m$, $h$, and $n$ are qualitatively related to the gating probabilities for activation and inactivation of the sodium channels and activation of the potassium channels, respectively, which characterize the opening and closing dynamics of the ion gates.

The total synaptic transmission between the neurons, following the ExpSyn model \cite{The NEURON Book}, is determined as the excitatory and inhibitory ion channels, where an input signal for the $\textrm{j}$-th pre-synaptic neuron is received by the $\textrm{i}$-th post-synaptic neuron,
\begin{equation*}
I^{(\textrm{i})}_{\textrm{syn}} = \sum_{\textrm{j} \in \Omega(\textrm{i})} G^{(\textrm{i},\textrm{j})}_{\textrm{syn}} \,\, ( V^{(\textrm{i})}_{\!\textrm{m}} - E_{\textrm{syn}}),
\end{equation*}
in which $G^{(\textrm{i},\textrm{j})}_{\textrm{syn}}$ describes the summed synaptic conductance of the neuron $\textrm{i}$ induced by its neighbors $\Omega(\textrm{i})$ and $E_{\textrm{syn}}$ denotes the synaptic reversal potential with the values of $0$ and $-70$ \si{\milli\volt} \cite{Schutter(1994),Neuroscience Book} to ensure the excitatory and inhibitory synapses, respectively. The chemical conductivity $G^{(\textrm{i},\textrm{j})}_{\textrm{syn}}$ itself takes a smooth exponential decay of the form
\begin{equation*}
G^{(\textrm{i},\textrm{j})}_{\textrm{syn}} (t) = w^{(\textrm{i},\textrm{j})}_{\textrm{syn}} \exp(- \dfrac{t}{\tau_{\textrm{syn}}}),
\end{equation*}
where $w^{(\textrm{i},\textrm{j})}_{\textrm{syn}}$ represents the coupling matrix of the synaptic weights between the neurons $\textrm{i}$ and $\textrm{j}$, and $\tau_{\textrm{syn}}$ is the synaptic decay time constant which sets $\tau_{\textrm{syn}} = 1.3$ and $\tau_{\textrm{syn}} = 0.3$ \si{\milli\second} \cite{Schutter(1994),Neuroscience Book} for the excitatory and inhibitory synapses, respectively.

In our study, regarding a desired relation between the sigmoidal response function used in the original Wilson-Cowan equations and the Gaussian input \cite{Wilson(1972),Zandt(2014)}, the external current $I^{(\textrm{i})}_{\textrm{stim}}$ is a norml random process with the mean parameter $\bar{I}_{\textrm{stim}}$ and the variance $\sigma_{I_{\textrm{stim}}}^{2}$, and we suppose that stimulus configurations involve constant inputs to only the excitatory subpopulation.

This approach captures the membrane voltage dynamics induced by the synaptic and injected currents at the level of soma. To improve computational efficiency and analytical tractability, somas are established in NEURON software \cite{Hines(1997)}; the ODE part of this nonlinear system is implemented using the backward Euler's discretization scheme with a time step of $\Delta = 0.1$ \si{\milli\second} for a total duration of $T = 10^{3}$ \si{\milli\second}. All parameter values for the soma and the Hodgkin-Huxley model which refer to the experimental observation \cite{Neuroscience Book,Johnston(1995)} are included in the Supplementary Material.
\subsection{Interconnection topology}

Physiological evidence suggests that the cortex of the human brain contains about $50$ billion neurons: $80\%$  are excitatory type, whereas the remaining $20\%$ are inhibitory type \cite{Sholl(1956),Cowan(2016)}. It is also reported that when assuming a Gaussian distribution of synapses or thresholds, the reasonable hypothesis results in the sigmoid activation of the Wilson-Cowan oscillators that is the integral of a Gaussian curve \cite{Wilson(1972)}. In this simulation, we consider a population of $N_{\textrm{exc}} = 0.8 N$ excitatory neurons and $N_{\textrm{inh}} = 0.2 N$ inhibitory neurons with conductance dynamics introduced in the present section. Furthermore, in order to confirm the tendency for the Wilson-Cowan restrictions, the probability distribution of the synaptic connections within the EI population follows a Gaussian, where the average and the standard deviation are reflected in the parameters $\mu_{\textrm{p-q}}$ and $\sigma_{\textrm{p-q}}^{2}$ with $\textrm{p-q} = \{ \textrm{E-E}, \textrm{E-I}, \textrm{I-E}, \textrm{I-I} \}$ denoting the excitatory and inhibitory interactions, but let the variation in the level of spiking threshold is negligible. Figure\,\hyperlink{Fig1}{1} illustrates a graphical representation for the network structure based on bidirectional communications. For the biophysical population model, we employ the global synaptic weight parameter $W$ to ensure that $w_{\textrm{syn}}$ is linearly proportional to $N$. Indeed, this may scale the couplinge weights so that the sum of all synaptic strengths of an individual neuron, without depending on the network size and the connectivity number, is equal to $W$. Thus, the global synaptic weight allows the large-scale characteristic of the system to be created randomly, given local Gaussian structure.
\subsection{Population firing rates}

The geometric form of the firing rate function, being a probability density, is very important; there exist many choices made for biological consistency that various authors utilize. The distribution of firing rates within the localized EI population in the Wilson-Cowan model is intuitively determined according to a general sigmoid nonlinearity \cite{Wilson(1972)}. To convert the spike trains of post-synaptic neurons into the pulse-like rate, we provide a Gaussian approximation \cite{Theoretical Neuroscience Book}, given dynamically equivalent sigmoidal response \cite{Zandt(2014)},
\begin{equation*}
Rate (t) = \dfrac{1}{\sqrt{2\pi}\,\sigma_{Rate}} \exp\!\Big(\!-\dfrac{(t - \mu_{Rate})^{2}}{2\,\sigma_{Rate}^{2}}\Big)
\end{equation*}
with the mean $\mu_{\textrm{Rate}}$ and the variance $\sigma_{Rate}^{2}$ that describes the average firing rate at time $t$ taking into account spikes generated both before and after $t$. In this expression, the standard deviation is based primarily on sustaining an approximately constant rate of activity in the absence of any neural coupling and the state of stochastic firing.
\section{ACKNOWLEDGMENTS}
\hypertarget{Ackn}{}

The third author is indebted to the Research Core: ``Bio-Mathematics with computational approach" of Tarbiat Modares University, with Grant Number ``IG-39706".
\pdfbookmark[section]{REFERENCES}{bibliography}

\end{document}